\documentclass[authoryear,review,12pt]{elsarticle}
\usepackage{graphicx,natbib,amssymb}
\journal{New Astronomy}

\newcommand{\kms}{\,km\,s$^{-1}$}

\begin{document}
\begin{frontmatter}
\title{V2653 Ophiuchii with a pulsating component and $P_{puls}$ -  $P_{orb}$, $P_{puls}$ - $g$ correlations for $\gamma$-Dor type pulsators}
 \author[l1]{\"{O}. \c{C}ak{\i}rl{\i} \corref{cer}}
  \author[l1]{ C.~Ibanoglu}
\address[l1]{Ege University, Science Faculty, Astronomy and Space Sciences Dept., 35100 Bornova, \.{I}zmir, Turkey \corref{cer}}
\cortext[cer]{Corresponding author. Tel.: +902323111740; Fax: +902323731403 \\
E-mail address: omur.cakirli@gmail.com}

\begin{abstract}
We present new spectroscopic observations of the double-lined eclipsing binary V2653\,Oph. The 
photometric observations obtained by $ASAS$ were analysed and combined with the analysis of radial 
velocities for deriving the absolute parameters of the components. Masses and radii were determined for 
the first time as M$_p$=1.537$\pm$0.021 M$_{\odot}$ and R$_p$=2.215$\pm$0.055 R$_{\odot}$, M$_s$=1.273$\pm$0.019 M$_{\odot}$ 
and R$_s$=2.000$\pm$0.056 R$_{\odot}$ for the components of V2653\,Oph. We estimate an interstellar 
reddening of 0.15$\pm$0.08\,mag and a distance of 300$\pm$50\,pc for the system, both supporting the 
membership of the open cluster Collinder\,359. Using the out-of-eclipse photometric data we have made frequency 
analysis and detected a periodic signal at 1.0029$\pm$0.0019\,c/d. This frequency and the location of the
more massive star on the HR diagram lead to classification of a $\gamma$ Dor type variable. Up to date
only eleven $\gamma$ Dor type pulsators in the eclipsing binaries have been discovered. For six out of 11 systems,
the physical parameters were determined. Although a small sample, we find empirical relations that 
$P_{puls}$ $\propto$ $P_{orb}^{0.43}$ and $P_{puls}$ $\propto$ $g^{-0.83}$. While the pulsation periods 
increase with longer orbital periods, they decrease with increasing surface gravities of pulsating components 
and gravitational pull exerted by the companions. We present, briefly, the underlying physics behind 
the correlations we derived.
\end{abstract}
\begin{keyword}
Stars: fundamental parameters --
individual: V2653\,Oph -- 
binaries: eclipses --
                   --
\end{keyword}   
\end{frontmatter}
%
%
\section{INTRODUCTION }
\label{sec:intro}
Main-sequence and subgiant stars of spectral type A to F in detached and semi-detached eclipsing
binaries are very interesting targets for asteroseismic studies, i.e., the determination of the boundaries
of instability strip by interpreting the observed oscillation characteristics. Indeed,
eclipsing binaries containing A-F stars offer the possibility of determining accurate masses and radii 
for the components. Unique advantage is if the binary resides in a cluster, determination 
of the fundamental properties and, in particular for systems belonging to the open clusters, can 
be used to address relevant astrophysical issues, such as chemical compositions, age and the 
distance of two stars with accurate masses and radii, allowing a much more discriminating test of 
physical ingredients of theoretical models.

We present a study of the detached {\em Algol} type eclipsing binary, which is probably a member 
of the open cluster Collinder\,359 \citep{zejda}. We have obtained complete light curves
from the photometry obtained by {\em Hipparcos} \citep{Perryman+97aa} and All Sky Automated Survey, 
ASAS, \citep{Poj02}. We have also obtained an extensive spectroscopy using the FRESCO {\'e}chelle
spectrograph at the telescope of Catania Astrophysical Observatory.

\cite{och} measured the visual magnitude of 8.90 mag with a precision of one tenth of a magnitude
and a color of G5 using "microfiche" photographic observations for V2653\,Oph. The eclipsing character
of V2653\,Oph (HD\,162811;BD+03$^{o}$3514; HIP\,87511; V=9.49, B-V=0.60 mag) \footnote{adopted 
from the SIMBAD astronomical data base.} was discovered by the $Hipparcos$ satellite mission, the primary 
eclipse having an amplitude of 0.09\,mag. The depth of the secondary minimum is nearly equal to that of the 
primary one. The first eclipse light curve was roughly revealed by the $Hipparcos$ observations. \citet{Kaz99} 
designated it as V2653\,Oph in the {\sc 74th Name-list of Variable Stars}, and classified it as an eclipsing 
binary. In recent years, large-scale photometric surveys, such as All Sky Automated Survey (ASAS) 
\citep{pojmanski}, $MOST$\citep{wal}, $CoRot$\citep{bag}, $Kepler$ \citep{gil},etc.  have been conducted with the main aim of looking for 
transiting exoplanets. The valuable by-product of this search has been the very large number of well sampled 
eclipsing binary light curves. One of the eclipsing binaries observed at the Las Campanas Observatory as part 
of the ASAS was V2653\,Oph. Yet the astrophysical parameters are not obtained for the system.

\section{Observation and data reduction}
\label{sec:observation}
The 91-cm telescope of the Istituto Nazionale di Astrofisica -- Osservatorio Astrofisico di Catania (INAF--OACT) 
was used to carry out spectroscopy of the target. Details of the telescope and the echelle spectrograph are given
by \citet{or}. Observations were carried out during eleven nights between 21 June - 05 July 2008. Exposure
times were fixed for target star at 2400\,s. 
 
The reduction of the spectra was done by using the NOAO/IRAF package\footnote{IRAF is distributed by the 
National Optical Astronomy Observatory, which is operated by the Association of Universities for Research in 
Astronomy, Inc.}. The {\em Th-Ar} emission line spectra were used to calibrate the wavelength references 
of each observations. The S/N ratio of the spectra was at least $\sim$\,90.

Photometric observations of the system obtained and published in ASAS-3 were used to construct the V-band light curve.
For calculation of the orbital phases we tried to improve a better epoch and an orbital period for the system.
An orbital period of about 4.394 days was estimated from the photometric observations of the system by 
the $Hipparcos$ mission. We searched for orbital period using data combined from 
the {\em Hipparcos} and ASAS surveys. Analysis of Variance (AoV) algorithm
developed by \citet{AoV89} and implemented in the VarTools light curve analysis program \citep{hartman08} was
used for detection of sharp periodic signals. We obtained the period and {\em False Alarm Probability} of peaks which
are significant at a level greater than 3$\sigma$. We detect significant peaks in the power spectra
at periods between 2--5 days, the most significant being at 4.39\,days. We improved the light elements of the system
using the Wilson-Devinney code \citep{Wil03}. The light and radial velocity curve of the system is phased with 
the following ephemeris and period:

\begin{equation}
 T_{\mathrm{minI}} = 2453189.668(58) + 4^d.3942894(48)  \times E.
\end{equation}

The orbital period is very close to that estimated by \citet{ote}. The radial velocities and photometric observations
were phased using these light elements. 

\section{Spectroscopic data}
\subsection{Radial velocity analysis}
\label{sec:rv}
Radial velocities (RV) of the components were derived with a standard cross correlation algorithm, IRAF's tool FXCOR. We
used the wavelength interval 4500 -- 5000  \AA, which is rich in metallic lines. As cross correlation template, we 
chose the nearby primary target $\iota$\,Psc (F7\,V) of similar spectral type and we used, for reference, as the RV 
standard. The radial velocities of the components were derived by fitting two Gaussian curves with the FXCOR 
function and presented in Table\,1. The radial velocities were analysed using the {\sc RVSIM} software program 
\citep{Kan07}. The parameters are presented in Table\,2 and the fits are compared with the observations in Fig.\,1.

\begin{table}
\centering
\caption{Heliocentric radial velocities of V2653\,Oph. The columns give the heliocentric 
Julian date, orbital phase and the radial velocities of the two components with the corresponding 
standard deviations.}
\begin{tabular}{ccrcrcc}
\hline
HJD--2400000 & Orbital& \multicolumn{4}{c}{V2653\,Oph}&  	\\
             & Phase  & $V_{\rm P}$                   & $\sigma$   & $V_{\rm S}$ & $\sigma$	\\
\hline
54638.34598 &	  0.6729  &   84.2&  0.6& -83.2 &   0.7  \\
54639.33703 &	  0.8984  &   44.2&  0.9& -55.4 &   0.7  \\
54640.33176 &	  0.1248  &  -81.2&  1.1&  55.6 &   1.1  \\
54641.34108 &	  0.3545  &  -81.2&  0.9&  51.4 &   0.4  \\
54642.34823 &	  0.5837  &   50.1&  1.1& -55.3 &   1.1  \\
54643.35360 &	  0.8124  &   81.5&  1.0& -84.3 &   0.6  \\
54644.38749 &	  0.0477  &  -41.3&  1.4&  21.2 &   0.7  \\
54647.37993 &	  0.7287  &   92.4&  0.7& -88.5 &   0.7  \\
54649.36320 &	  0.1800  & -100.2&  0.8&  66.5 &   1.2  \\
54651.35779 &	  0.6339  &   71.4&  0.8& -71.2 &   1.3  \\
54654.34120 &	  0.3129 &  -95.6&  0.7&  66.2 &   1.1  \\
\hline
\end{tabular}
\end{table}
\begin{table}
\centering
\caption {Results of the radial velocity analysis for V2653\,Oph.}
\begin{tabular}{@{}lcccccccc@{}c}
\hline
Parameter  & \multicolumn{2}{c}{V2653\,Oph}&  	\\
             & Primary           & Secondary   	\\
\hline
$K$ (km\,s$^{-1}$) 			      &$ 83\pm$1	      			              &$93\pm$1	      \\
$V_\gamma$ (km\,s$^{-1}$) 		&\multicolumn{2}{c}{$-8.41\pm$0.43}  			            \\  
Average O-C (km\,s$^{-1}$)		& 0.5         	      			          &0.9      	    \\
$a\sin i$ ($R_{\odot}$)			  &\multicolumn{2}{c}{$15.9\pm$0.1}  			              \\
$M\sin^3i$ ($M_{\odot}$)		  & 1.50$\pm$0.10      			            & 1.25$\pm$0.18	\\      
$e$					                  &\multicolumn{2}{c}{0.0}  			                      \\
\hline
\end{tabular}
\end{table}

\subsection{Determination of the atmospheric parameters}
\label{sec:class}
The width of the cross-correlation function (CCF) is a good tool for the measurement of 
$v\sin i$ of a star. We used a method developed by \citet{Pen96} 
to estimate the $v\sin i$ of each star composing the investigated double--lined eclipsing binary
system from its CCF peak by a proper calibration based on a spectrum of a narrow-lined star
with a similar spectral type. The rotational velocities of the components were obtained by 
measuring the FWHM of the CCF peak related to each component in five high-S/N spectra acquired 
near the quadratures, where the two spectra have the largest relative difference in radial 
velocity. The CCFs were used for the determination of $v\,sin\,i$ through a calibration of the 
full-width at half maximum (FWHM) of the CCF peak as a function of the $v\,sin\,i$ of artificially 
broadened spectra of a slowly rotating standard star ($\iota$\,Psc, $v\sin i \simeq$3\,km\,s$^{-1}$, 
e.g., \citealt{takeda}) acquired with the same set up and in the same observing night as the target 
system. The limb darkening coefficient was fixed at the theoretically predicted values, 0.55 for 
the system \citep{Van93}. We calibrated the relationship between the CCF Gaussian width and 
$v\,sin\,i$ using the \citet{Con77} data sample. This analysis yielded projected rotational 
velocities for the components of V2653\,Oph as $v_{\rm P}\sin i$=25~km\,s$^{-1}$, and 
$v_{\rm S}\sin i$=20~km\,s$^{-1}$. The mean deviations were 1 and 2 km\,s$^{-1}$, for the 
primary and secondary, respectively, between the measured velocities for different lines. 

In Fig.\,2 we show the simulation of the spectrum of the system with that of broadened combination
of the two standard spectra at phase 0.72 (HJD 54647.37993) using the derived rotational velocities.
The spectral order at blue wavelength is displayed. The sum of residuals is also shown, as a function
of $v\,sin\,i$, in the insets of Fig\,2.

\begin{table}
\centering
\begin{minipage}{85mm}
\caption {Spectral types, effective temperatures, surface gravities, metallicity and 
rotational velocities of components estimated from the spectra of V2653\,Oph.}
\begin{tabular}{@{}lcccccccc@{}c}
\hline
Parameter  & \multicolumn{2}{c}{V2653\,Oph }&  					                      \\
		    & Primary           & Secondary   					                          \\
\hline
Spectral type 				        & F(2$\pm$0.5)\,V	        &F(7$\pm$0.5)\,V	    \\
 $T_{\rm eff}$ (K)	    		  &6\,950$\pm$480  	        &6\,350$\pm$650		    \\   
 $\log~g$ ($cgs$)			        & 4.07$\pm$0.02           &4.22$\pm$0.06       	\\    
 $v\sin i$ (km\,s$^{-1}$)  		&25$\pm$1	  	            & 20$\pm$2      	    \\
 $[Fe/H]$ (dex)  			        &-0.11$\pm$0.08	  	      &0.07$\pm$0.09        \\   
 \hline
\end{tabular}
\end{minipage}
\end{table}

We also performed a spectral classification for the components of the system using the COMPO2, an 
IDL\footnote{Interactive Data Language, ITT 1997} code for the analysis of high-resolution spectra of eclipsing binary 
systems originally written by \citet{Frasca2006} and adapted to the {\sc REOSC} spectra. This code searches for 
the best combination of two reference spectra able to reproduce the observed spectrum of the system. We 
give, as input parameters, the radial velocities and projected rotational velocities of the two 
components of the system, which were already derived. The code then finds, for the selected spectral 
region, the spectral types and fractional flux contributions that best reproduce the observed 
spectrum, i.e. which minimize the residuals in the collection of difference (observed\,$-$\,composite) 
spectra. For this task we used reference spectra taken from the \citet{Val04} $Indo-U.S.\ Library\ of\ 
Coude\ Feed\ Stellar\ Spectra$ (with a a resolution of $\approx$\,1\AA) that are representative of stars 
with spectral types from late-O type to early-M, and luminosity classes V, IV, and III. The atmospheric 
parameters of these reference stars were recently revised by \citet{Wu2011}.

We selected 198 reference spectra spanning the ranges of expected atmospheric parameters, which means 
that we have searched for the best combination of spectra among 39\,204 possibilities per spectrum. The 
observed spectra of V2653\,Oph in the $\lambda$4960--5000 spectral region were best represented by the 
combination of HD\,183085 (F0.5\,V) and $\iota$ Psc (F7\,V). However, we have adopted, for each component, 
the spectral type and luminosity class with the highest score in the collection of the best combinations 
of templates, where the score takes into account the goodness of the fit expressed by the minimum of the 
residuals. We have thus derived a spectral types for the primary and secondary component of V2653\,Oph 
as F2 and F7 main-sequence stars, with an uncertainty of about 0.5 spectral subclass. The effective 
temperature and surface gravity of the two components of the system are obtained as the weighted average 
of the values of the best spectra at phases near to the quadratures combinations of templates adopting 
a weight $w_i=1/\sigma_i^2$, where $\sigma_i$ is the average of residuals for the $i$-th combination. The 
standard error of the weighted mean was adopted for the atmospheric parameters. The atmospheric parameters 
obtained by the code and their standard errors are reported in Table\,3. The observed spectra of V2653\,Oph 
at phase near to the quadrature is shown in Fig.\,2 together with the combination of two reference 
spectra which gives the best match and their residuals.

\subsection{Spectroscopic light ratio}
For the detached eclipsing binaries, where the temperature of the two stars are closely similar, the 
light ratio can be calculated from the cross--correlation functions (CCF) of the various spectra. \cite{or}
showed, by measuring the average fitted full width half maximum ({\sc FWHM}) values of the echelle orders 
in the selected spectral domains, which include several photometric absorption lines. In this study we have 
disregarded very broad lines, such as H$_{\beta}$ and H$_{\alpha}$, because their broad wings affect the 
CCF and lead to large errors. Inspection of the CCFs showed that the two components are visible and well 
separated from each other near the quadratures. Following the method proposed by \cite{or}, the two-Gaussian 
fits of the well--separated CCFs using the de-blending procedure in the IRAF routine SPLOT. The average 
fitted FWHMs are 120$\pm$2 and 65$\pm$3 \kms for the primary and secondary components. Fig.\,3 shows a 
sample of the double-Gaussian fit. Indeed, the shapes and velocities corresponding to the peaks of the 
CCFs are slightly changed. We estimated the intensity ratio $\ell_1$ /($\ell_{1}+\ell_{2}$)=0.649$\pm$0.032 
from the two-Gaussian fits of the well-separated CCFs.

\section{Light curve analysis}
\label{sec:lc}
The photometric data collected from the ASAS catalogue. We analyzed the light curve using the 
Wilson-Devinney code \citep[hereafter WD; e.g.,][]{Wil71,Wil79,Wil03,Wil06} implemented into the software 
{\sc phoebe} \citep{Prs05}. The WD code is widely used for determination of the orbital parameters of the 
eclipsing binaries. To run the code we need some initial parameters. The initial logarithmic limb-darkening 
coefficients were taken from the tables given by \citet{Van93} as $x_1$=0.47 and $x_2$=0.48, and are automatically
interpolated at each iteration by {\sc phoebe}. The effective temperature of the 
primary star is taken as 6\,950 K and the ratio of the masses of two components $q$=0.84. We have started with 
$Mode-2$ meant for detached binary systems, keeping the temperature of the primary and the mass-ratio as fixed 
parameters. According to the WD code we  adjusted the following parameters: $i$ (the orbital inclination), 
$\Omega_1$ (the potential for the primary), $\Omega_2$ (the potential for the secondary), $T_{\rm eff_2}$ (the 
effective temperature of the cool star), L$_1$ (the luminosity of the primary), and the zero-epoch offset. The 
luminosity of the secondary star, L$_2$,  was constrained by the model. After a few numbers of runs of the Differential
Correction program in $Mode-2$ the sum of residuals squared showed a minimum and the corrections to the adjustable parameters 
became smaller than their probable errors. The results are presented in Table\,4. The corresponding computed
light curve is shown in Fig.\,4 as a continuous line.

\begin{table}
\centering
\begin{minipage}{85mm}
\caption{Final solution parameters for the detached model of V2653\,Oph. }
\begin{tabular}{@{}ccccc}
\hline
Parameters  &  V2653\,Oph 	\\                  
\hline
$i^{o}$			                    	    &81.78$\pm$0.25 	        \\
$T_{\rm eff_1}$ (K) 				          &6\,950[Fix]		          \\
$T_{\rm eff_2}$ (K) 				          &6\,540$\pm$1\,60	        \\
$\Omega_1$ 	          			          &8.048$\pm$0.175 	        \\
$\Omega_2$					                  &7.717$\pm$0.183 		      \\
$r_1$				                          &0.1391$\pm$0.0034	      \\
$r_2$		           	            	    &0.1256$\pm$0.0035	      \\
$\frac{L_{1}}{(L_{1}+L_{2})}$   		  &0.613$\pm$0.015	        \\
$\sum(O-C)^{2}$			            	    &0.188  		              \\	 
\hline  
\end{tabular}
\end{minipage}
\end{table}

\section{Absolute dimensions}
Combining the results of radial velocities and light curve analysis we have calculated the absolute 
parameters of the stars. The separation between the components of eclipsing pair is calculated as 
$a$=15.98$\pm$0.07 R$_{\odot}$. The fundamental stellar parameters for the components such as masses, 
radii, luminosities are listed in Table\,5 together with their formal standard deviations. The standard 
deviations of the parameters have been determined by using the {\tt JKTABSDIM\footnote{This can be obtained 
from http://http://www.astro.keele.ac.uk/$\sim$jkt/codes.html}} code, which calculates distance and other 
physical parameters using several different sources of bolometric corrections \citep{sou05}. The masses for 
the primary and secondary star are consistent with F2$\pm$0.5V and F7$\pm$0.5V stars.

\begin{table}
  \caption{Absolute properties of the V2653\,Oph components}
  \label{parameters}
  \begin{tabular}{lcc}
  \hline
   Parameter 						                    & Primary	              &	Secondary		        \\   
   \hline 
	Mass (M$_{\odot}$)				                & 1.537$\pm$0.021		    & 1.273$\pm$0.019		  \\
	Radius (R$_{\odot}$)				              & 2.215$\pm$0.055		    & 2.000$\pm$0.056		  \\   
	$T_{eff}$ (K)					                    &  6\,950$\pm$480	      & 6\,350$\pm$650	    \\
   $\log~(L/L_{\odot})$				              & 1.014$\pm$0.122	      & 0.820$\pm$0.175	    \\
   $\log~g$ ($cgs$) 					              & 3.934$\pm$0.021 	    & 3.941$\pm$0.024	    \\
   $Sp.Type$ 						                    & 	F(2$\pm$0.5)\,V		  &  F(7$\pm$0.5)\,V 		\\
   $M_{bol}$ (mag)					                & 2.21$\pm$0.31		      & 2.69 $\pm$0.44		  \\
   $BC$ (mag)						                    &   0.030               & 0.009    			      \\  
   $M_{V}$ (mag)						                &   2.18$\pm$0.33		    & 2.69$\pm$0.48	      \\  
   $(vsin~i)_{calc.}$ (km s$^{-1}$)	        & 25.5$\pm$0.6		      & 23.0$\pm$0.7			  \\       
   $(vsin~i)_{obs.}$ (km s$^{-1}$)	        & 25$\pm$1		          & 20$\pm$2 		        \\    
   $d$ (pc)							                    &   284$\pm$45  	 	    &  321$\pm$73  		    \\
\hline  
  \end{tabular}
\end{table}

The distance determination depends, of course, on the total apparent visual magnitude and the observed
colour of the system. The observed visual magnitude of the system was estimated by various authors
in the range from 9.48 \citep{mendo} to 9.98\,mag. \citep{hanson}. 
Since the Tycho measurements include probable errors, e.g. $B_{T}$=10.224$\pm$0.030, $V_{T}$=9.568$\pm$0.024 
mag,  we adopt the apparent visual magnitude of 9.509$\pm$0.027\,mag and colour index of (B-V)=0.558$\pm$0.082\,mag 
for the system, transformed from Tycho to Johnson photometric system using the coefficients given by \citet{hog}. Taking 
into account the de-reddened colour index of the primary star as 0.35 mag for an F2V	star and the light ratio of 
L$_2$ /L$_1$=0.63 we estimated de-reddened colour index for the system as 0.405 mag. Then we estimated the  
interstellar reddening of $E$(B-V)=0.15$\pm$0.08\,mag for the system.  The bolometric 
corrections are adopted from \citet{flo} for the primary and secondary stars. The absolute bolometric magnitude 
of the Sun is taken as 4.74. The distance to the system is estimated as 300$\pm$50\,pc. The weighted mean distance 
calculated from the individual distances given in Table\,5 is 298$\pm$56\,pc, consistent with the distance estimated 
from total apparent visual magnitude and interstellar reddening of the system. As it is known that the distance
measured by the $Hipparcos$ becomes inaccurate for d$>$100 pc.

V2653\,Oph is located, on the sky, very close to the centre of the Collinder\,359 open cluster. Recently \cite{zejda}
presented the first new catalog of known variable stars in open cluster regions. They have proposed that V2653\,Oph 
is a probable member of the  Collinder\,359. However, its membership to the Collinder\,359 is still uncertain. The 
systemic velocity of the V2653\,Oph, -8.4 \kms, is consistent with the measured cluster systemic velocities of \citet{plasket}, 
\citet{nord} and \citet{gon}. Interstellar reddenings for the stars in the cluster were estimated as 
of 0.16\,mag obtained by \citet{kha} and 0.191\,mag \citet{zejda}. The distance of the Collinder\,359 is still uncertain.
Distance determination is in general based on isochrone fitting of early type stars. For Collinder\,359 listed
13 cluster members by \cite{col} and estimated a distance to the cluster ranging from 210\,pc to 290\,pc. Later,
\cite{ruc} made UBVRI photometry and obtained a distance to the cluster about 436\,pc, rejecting half of the original 
members. Trigonometric parallaxes of five photometric members were measured by $Hipparcos$ satellite which 
yielded distances between 260\,pc and 280\,pc for Collinder\,359. The systemic velocity, interstellar reddening 
and distance of  V2653\,Oph are in good agreement with those derived for the cluster. It seems highly probable 
that V2653\,Oph does belong to Collinder\,359. 

The age determined for the Collinder\,359 has changed dramatically since the early determination by \cite{wi} of 
20 to 50\,Myr. Later studies derived 30\,Myr \citep{abt}. Both results are in agreement and consistent with the 
more recent values range from 50\,Myr \citep{ba,kha}. \cite{jef} put an upper limit of 100\,Myr on the age of 
Collinder 359 using the lithium test described by \cite{reb}. Comparison with the various evolutionary tracks 
indicates that the components of V2653\,Oph have an age of about 300($\pm$100)\,Myr, older than the cluster. If the 
binary is really a member of the cluster its age should be at least three times older than the values estimated 
up to date.

\section{$\gamma$ Dor type variables in close binary systems}
Fig.\,4 shows that there is a considerable light variations at outside of eclipses. The $\gamma$ Dor type
pulsating single stars and components in the eclipsing binaries were plotted in the log\,T$_{eff}$-log\,L/L$_{\odot}$
plane and shown in Fig.\,5. The empty symbols show the single stars taken from the catalog of \citet{henry} while 
the filled circles indicate the $\gamma$ Doradus type pulsators in the eclipsing binaries. The primary component 
of V2653\,Oph locates in the instability strip of $\gamma$ Doradus type variable stars.  The known $\gamma$ Dor 
type pulsating variables in the eclipsing variables are: V551\,Aur, VZ\,CVn, V2094\,Cyg, V404\,Lyr,
Corot\,100866999, Corot\,102918586, Corot\,102937335, Corot\,102980178,KIC\,04544587, KIC\,11285625. 

The $\gamma$ Doradus and $\delta$ Scuti type pulsators share a similar parameter space in the Hertzsprung-Russell (HR) diagram.
While the $\delta$ Sct stars are believed to be mostly low-radial-order pressure mode pulsators \citep{bre00}, the $\gamma$ Dor
stars are high-radial-order gravity mode pulsators \citep{kaye}. \cite{handler02} showed that the $\delta$ Sct and  $\gamma$ Dor 
type pulsating stars are separated by the values of their pulsation constants. While the $\delta$ Sct pulsators have pulsation 
constant of $Q\sim0.03$ days, the known $\gamma$ Dor pulsators all have $Q$ $>$ 0.23 days.
      
Therefore, we started to search any periodic light variations in the system. To investigate the intrinsic light
variations of the components in eclipsing binary systems, the best-fit binary star model was subtracted from the
light curve. This leaves the stellar pulsations, random noise, systematic calibration noise, and any slight
mismatches between the model and actual binary star light curve. In these circumstances, frequency analysis
was performed on the residuals of the system, in the region 0--3 d$^{-1}$, where the most 
of the dominant pulsation frequencies are found. In order to generate the frequency spectrum of residual data, shown in 
Fig.\,6, we used {\tt Period04}\footnote{https://www.univie.ac.at/tops/Period04/}. Figure\,6 displays the 
amplitude spectra in the frequency range from 0.6 to 2.0 d$^{-1}$, $\gamma$ Dor oscillation 
frequencies indeed are typically detected in this range \citep{balo}.

The amplitude spectra after the analysis are presented in Fig.\,6. Clearly 
visible is the only a peak around 1.0 d$^{-1}$. The light variation is dominated with a frequency 
$f_1$=1.0029(19) and a full amplitude of about 60 mmag. The uncertainty of the frequency was derived according
to \cite{kallinger}. No significant peaks are present in the region higher than 5 d$^{-1}$ and smaller than 
0.6 d$^{-1}$. One sees that V2653\,Oph pulsates in the frequency range of $\gamma$ Dor type variables \citep{gri}.

Investigation of $\gamma$-Dor type pulsating variables in close eclipsing binary systems is a 
relatively new area of the stellar astrophysics. The first $\gamma$ Dor type pulsation was discovered
by \cite{iban} in the more massive primary star of the eclipsing binary system VZ\,CVn. Following this
discovery new $\gamma$-Dor pulsators were revealed by precise photometric studies. We collected the
eclipsing binaries with $\gamma$-Dor pulsators and presented in Table\,6. So far, 11 eclipsing binary systems 
containing a $\gamma$-Dor type pulsating component have been discovered. Unfortunately, spectroscopic observations
of only six systems have been made and precise radial velocities and, therefore, the orbital elements were determined.
The number of $\gamma$-Dor type pulsators in the close binaries will be rapidly increased in a few years after
the analyses of the huge precise data obtained by space missions such as $CoRot$, $Kepler$, etc.

A decade ago \cite{soy} suggested a linear relationship between the orbital and pulsation periods for
the 20 $\delta$-Sct type pulsators in the classic Algol type systems. This relationship based on observations
is confirmed and refined by \cite{lia} using the data almost quadruple than that of \cite{soy}. The correlation 
between the pulsation ($P_{puls}$) and orbital ($P_{orb}$) periods has been theoretically established by \cite{zhang}.
In Table\,6 we present mass and radius (in solar units), mean density and surface gravity (in CGS units) and
pulsation constant (in days) for the pulsating stars
in the eclipsing binaries, for which precise photometric and spectroscopic observations are available.
The dominant pulsation periods are taken into account and the pulsation constants were calculated using these 
periods and mean densities. The pulsation constants vary between 0.17 and 0.72, systematically larger than the
value of 0.033 days for the radial fundamental modes \citep{bre,sta}. 
      
In Fig.\,7 we plot the dominate pulsation periods against the orbital ones, where the periods are in days. Even 
with a small sample, 11 systems at present, we claimed sufficient evidence of a correlation. One possible 
explanation would be that according to this empirical correlation, the longer the orbital period of a binary, the 
longer the pulsation period of its pulsating primary.  The $P_{puls}-P_{orb}$ relation appears to be a simple 
linear form, the following numerical relation was obtained by linear least squares method, 

\begin{equation}
log P_{puls} = 0.425(\pm 0.016) \times \log P_{orb} - 0.355(\pm 0.087).      
\end{equation}
The correlation coefficient for this relationship is about 0.79, indicating its significance. The scatter appears to 
increase towards the shorter orbital periods.

Although the sample contains eleven pulsating primary stars and existence of a considerable scatter in the observed
pulsation frequencies a connection between pulsation and orbital periods seems to exist in the $\gamma$-Dor type stars
as pointed out by \cite{soy} for the $\delta$-Sct stars. The pulsation periods are proportional to the orbital periods as 
$P^{0.43}$. This relation indicates that the shorter orbital period the smaller pulsation period. For the 
$\delta$-Sct stars in close binaries \cite{lia} find a relationship as $P_{puls}$ proportional to $P_{orb}^{0.58}$.

Recently \cite{zhang} indicated that the upper limit of $P_{puls}/P_{orb}$ is about 0.09 for the $\delta$-Sct stars.
All the stars given in Table\,6 have larger ratios than those of the $\delta$-Sct stars except CoRot102980178. It has the 
smallest ratio with 0.072. This star is not clearly classified yet. It was classified as a $\gamma$-Dor pulsator by \cite{sok}, 
however, the classification as a {\tt Slowly Pulsating B star} (SPB) could not be ruled out. There are some doubts about the 
classifications of V551 Aur and KIC04544587. Both the Q-values and period ratios lead to classification of $\gamma$-Dor type 
pulsation. The pulsational period is connected to the mean density in the pulsating variables. The period-mean density 
relationship can be written as: 

\begin{equation}
P_{puls}= Q \sqrt{\rho_{\odot}/\rho}                          
\end{equation}
where Q is the pulsation constant, $\rho$ and $\rho_{\odot}$ are the mean density of the star and the sun. The mean 
density is expressed by the mass and radius of the star. Then the above equation can be rearranged to the form:

\begin{equation}
P_{puls}= Q \sqrt{ \frac{R_{puls}^3}{M_{puls}} }
\end{equation} 
where R$_{puls}$ and M$_{puls}$ are the radius and mass of the pulsating star. Solving the Kepler's Third 
law for the mass of the pulsator and inserting to the above equation we obtain,

\begin{equation}
P_{puls}= Q \times P_{orb} \times \bigg\{G \times \frac{R_{puls}^3} {4\pi^2 \times a^3}\bigg\}^{\frac{1}{2}} 
\times \bigg\{1-\frac{M_s}{M_{puls} + M_{s}}\bigg\}^{-\frac{1}{2}}
\end{equation} 
where P$_{orb}$ is the orbital period, $a$ is the semi-major axis of the eclipsing system's orbit and M$_{s}$) is the mass of 
the secondary star and G is the Newton's gravitational constant. This expression indicates that the pulsational 
period of a pulsator in an eclipsing binary is depended up on the orbital period. 

We also plot the pulsation periods against the gravities of the pulsating stars in Fig.\,8, where the pulsational periods 
are in days and the gravities in CGS units. The open square in this figure shows V404\,Lyr for which no radial velocities 
exist, the gravity is estimated from the preliminary mass and radius determined only from the light curves. Therefore it 
was not used in the calculation of the correlation between the pulsational period and gravity. Gravities were calculated 
for only six $\gamma$-Dor type pulsators in the close binaries for which radial velocities and photometric light curves 
exist. A preliminary least squares analysis yields a relationship between the pulsational period and gravity as,     

\begin{equation}
log P_{puls} = -0.828(\pm 0.076) \times \log g + 3.385(\pm 0.218)
\end{equation}
with a correlation coefficient of 0.81. The pulsational periods are decreasing while the surface gravities are increasing, similar 
to the $\delta$-Sct stars (dot-dashed line in Fig.\,8) \citep{lia} and to those all radially pulsating variable stars (dashed line) 
\citep{fer}. While the slope is about -3.3 for the $\delta$-Sct stars, -0.88 for the radially pulsating variable stars. The pulsation 
periods of $\gamma$-Dor type pulsators seem to depend on the surface gravity similar to those radially pulsating variable 
stars proposed by \cite{fer}.

The gravitational force applied to per gram matter on the surface of the pulsating component can be determined by
 
\begin{equation}
g =  G \times \frac{M_{puls}} {R_{puls}^2}.                           
\end{equation} 
Since P$_{puls}$ depends on the R$_{puls}$ and M$_{puls}$ as given in Eq.\,4 we can easily obtain,

\begin{equation}
P_{puls} = Q \times G^{\frac{1}{2}} \times R_{puls}^{ \frac{1}{2}} \times g^{-\frac{1}{2}}
\end{equation}
The pulsational period is inversely proportional of the square root of the surface gravity of the pulsating star.

We also plot in Fig.\,9 the pulsational periods of the oscillating primary components against the gravitational 
pull exerted to the per gram of the matter on the surface of the pulsators by the less massive companions 
(in cgs units). The short orbital period binary VZ CVn appears to have longer pulsational period with respect to 
the gravitational pull of the companion. This star is significantly separated from the other stars. It has the 
greatest filling factor with 0.78 amongst the six stars (see  Table\,6). Excluding VZ CVn, the linear least-squares 
fit gives the following relationship,  
 
\begin{equation}
log P_{puls} = -2.021(\pm 0.012) \times \log (F/M_{puls}) + 2.093(\pm 0.098)      
\end{equation} 
with a correlation coefficient of 0.83. The high-value of the correlation coefficient indicates that this empirical 
relationship is very significant. As the force per unit mass on the surface of the pulsating star applied by the 
companion star is increased the pulsational period is decreased. The force exerted to per gram matter on the 
surface of the pulsating star can be calculated as,

\begin{equation}
\frac{F} {M_{puls}} = \frac{GM_{s}} {(a-R_{puls})^2}.
\end{equation} 
Since $P_{puls}$ $\propto$ $R_{puls}^{3/2}$ we can obtain $P_{puls}$ $\propto$ $(\frac{F}{M_{puls}})^{-3/4}$.
As the gravitational pull exerted on the per gram matter of the pulsator is increased the pulsational period is 
decreased, confirming the empirical relationship found.

Even with a small sample, pulsation periods of the $\gamma$-Dor type pulsators in the close binaries seem to
depend on the orbital periods of the systems, to the surface gravities of the pulsating primaries, and also 
to the gravitational pull applied by their companions. These empirical relationships should, of course, be 
checked by increasing the number of the sample.

Although we are still at the beginning of understanding the whole extent of the $\gamma$ Doradus 
phenomenon, but the combination of the current observational results and theoretical models are 
expected to answer a lot of open questions. Verification of the present results, especially for 
the low amplitude frequencies, the detection of the multi-frequencies, and the derivation of the 
excitation modes are some open questions on the class of $\gamma$ Dor type asteroseismology. We 
strongly suggest that accurate photometric survey of these systems are needed for answering the 
listed open questions. Moreover, phase coverage of their light curves and radial velocities will 
provide their geometrical characteristics. Finally, high-resolution spectroscopy are also suggested 
not only for the pulsations, but also for the components' radial velocity curves, which will lead 
us to calculate their absolute properties with high accuracy.

\section{Conclusion}
V2653\,Oph is a detached eclipsing binary in the young open cluster Collinder\,359. It is composed of main 
sequence F type stars. We have derived the astrophysical parameters of the components and other 
properties for the first time, listed in Table\,5, by combining the results obtained from the analyses 
of light curve and radial velocities. The masses and radii of the stars were derived with accuracy of 
2 and 6 percent, respectively. Both components are well inside their corresponding Roche lobes. Both 
components show nearly synchronized rotation. Taking into account the total apparent visual magnitude 
of the system, the light contribution of the stars and the bolometric corrections given by \cite{flo} and \cite{gir} 
we calculated a mean distance to V2653\,Oph as 300$\pm$50\,pc. The published parallax, measured by 
$Hipparcos$ satellite is very uncertain, $\pi$=3.72$\pm$1.28\,mas, corresponding to a formal distance of 
269$_{-68}^{+141}$ pc. The distance we derived is in good agreement with that of $Hipparcos$, but is 
more accurately determined. Both the interstellar reddening of 0.15$\pm$0.08\,mag and a distance of 
300\,pc are supporting the membership of the Collinder\,359.

After subtracting the binary model the residual light curve is obtained and analized for additional periodic signals. This 
analysis shows pulsation with a dominant frequency of about $f_1$=1.0029$\pm$19\,c/d, corresponding to 
0.9971$\pm$0.0019 d. We calculate the pulsation constant as 0.479 days. The spectral type of the primary star, the 
dominant pulsation period and pulsation constant suggest that the more massive primary star is most probably a 
$\gamma$ Doradus type pulsator. As it is shown in Fig.\,5 the pulsating primary star locates well inside the 
$\gamma$ Dor instability strip of the Hertzsprung-Russell diagram. The primary star seems to pulsate in 
multiperiodic modes. Further continuous multi-band photometric observations are needed for a better determination 
of the pulsational periods and amplitudes. 

We have suggested preliminary empirical relationships between the pulsation and orbital periods. In addition the 
pulsation periods of the primaries seem to correlate with their surface gravities and gravitational pull applied 
by their secondaries. We present some theoretical considerations which explain underlying physics behind these 
empirical relationships. Moreover, a physical interpretation of the empirical relations correlating the pulsation 
and orbital periods, and the location in the HR diagram with the evolutionary stage of the $\gamma$ Dor 
components in close binaries is certainly needed. Future theoretical pulsational mechanisms for pulsational 
behavior should take into account quantities such as mass, radius, radius expansion rate and dominant 
pulsation period. New observations of the binaries containing $\gamma$ Dor type components will increase 
the current example to better characterization of this type of asteroseismology.

\begin{table*}
\scriptsize
 \setlength{\tabcolsep}{2.5pt} 
  \caption{$\gamma$ Dor type pulsating stars in eclipsing binary systems. For each system, the colums 
2-13 are he orbital period, pulsational period (in days), masses of the primaries, fractional radii, radii in 
solar units, mean density in CGS, comparison with the Roche radii, pulsation constant (in days), mass of 
the secondary, orbital separation in solar units, gravitational force exerted by the secondary per gram 
matter of the primary and references}
  \label{parameters}
  \begin{tabular}{lccccccccccccr}
  \hline
System	                                    &P$_{orb}$ & P$_{puls}$ &$q$	      &	M$_p$/M$_{\odot}$	  &	 $r_p$,R$_{\odot}$	&	$\rho$	    &g, $\log$g &	$r{_p}/r_{Roche}$	   &	 $Q$	  & M$_s$  &	$a$(R$_{\odot}$)     & F/M$_p$ &   Ref	         \\
   \hline
V551\,Aur ($\delta$ Sct?)	                  &	1.1732	   &	0.12892	    &0.725	    &	  	                &	0.221	              &	0.7500	    &	          & 0.539         	     &		      &	       &        &         &   \cite{Liu}    \\
VZ\,CVn	                                    & 0.84246	   &  1.06876	    &0.778	    &1.83	                & 0.3115              &	0.4985	    & 16\,778   &0.777	               &  0.635   & 1.42	 & 5.554	& 2\,663	& \cite{iban}	    \\
                                            &            &              &           &                     & 1.730               &		          & 4.225     &	                     &          &        & 	      &	        &                 \\  
V2094\,Cyg	                                & 8.485376	 & 1.0995       & 0.4637	  & 1.79	              & 0.1024              &	0.1673	    & 8\,045    &0.229	               & 0.379    & 0.83   & 24.13	& 48.5	  & \cite{caki}     \\     
                                            &            & 1.0428       &           &                     & 2.471               &		          & 3.906     &	                     & 0.359    &        & 	      &	        &                 \\    
V404\,Lyr	                                  & 0.730943   & 0.50643      &0.385	    & 1.35	              & 0.4200              & 0.3500	    & 12\,023   &                      &          &        &        &         & 			\cite{Lee}\\
                                            &            & 4.47356      &           &                     & 1.76                &		          & 4.08      &	                     &          &        & 	      &	        &                 \\
V2653\,Oph                                  & 4.39429	   & 0.9971	      &0.792	    & 1.53	              & 0.1192              & 0.3248	    & 11\,879   &	0.299	               & 0.479	  & 1.292	 & 15.98	&  178	  & This paper      \\
                                            &            &              &           &                     & 1.88                &		          & 4.075     &	                     &          &        & 	      &	        &                 \\
CoRot100866999	                            & 2.80889	   & 0.62680      & 0.65		  &                     & 0.158			          &             &           & 0.378					       &          &        &        &         &\cite{cha}       \\
                                            &            & 0.73210      &           &                     &                     &		          &           &	                     &          &        & 	      &	        &                 \\
CoRot102918586                              & 4.39138    & 0.81665      & 0.898	    & 1.66                & 0.0992              & 0.5308      & 16\,936   & 0.256	               & 0.501    & 1.49   & 16.53  & 184     & \cite{mac}	    \\
                                            &            & 0.88758      &           &                     & 1.64                &		          & 4.229     &	                     & 0.545    &        & 	      &	        &                 \\
CoRot102937335                              & 3.97946    & 1.5772       & 0.3669    &                     & 0.1081              &             &           & 0.229                &          &        &        &         &	\cite{dam}      \\
                                            &            & 1.5748       &           &                     &                     &		          &           &	                     &          &        & 	      &	        &                 \\
CoRot102980178 ($\gamma$\,Dor or SPB?)	    & 5.0548	   & 0.3637	      & 0.206     &                     & 0.1068              &             &           &                      &          &        &        &         &		\cite{sok}	  \\
KIC04544587 ($\gamma$\,Dor or $\delta$\,Sct?)&2.189094   & 0.49721      & 0.808     & 1.98                & 0.1677              & 0.4633      & 16\,403   & 0.421                & 0.285    & 1.60   & 10.855 & 538     & \cite{ham}	    \\
                                            &            & 0.2883       &           &                     & 1.82                &		          & 4.215     &	                     & 0.165    &        & 	      &	        &                 \\
KIC11285625	                                & 10.79049	 & 1.7937       &0.778      &1.543                & 0.0737              & 0.2275      & 9\,376    & 0.184                & 0.720    & 1.200  & 28.8   & 46.6    & \cite{debos}	  \\
                                            &            & 1.7627       &           &                     & 2.123               &		          & 3.972     &	                     & 0.708    &        & 	      &	        &                 \\
\hline  \end{tabular} \end{table*}

\section*{Acknowledgments}
This study is supported by Turkish Scientific and Technology Council under project number 112T263.
The following internet-based resources were used in research for this paper: the NASA Astrophysics Data 
System; the SIMBAD database operated at CDS, Strasbourg, France. 

\newpage

\begin{figure}
       \includegraphics[width=9cm,height=13cm,keepaspectratio]{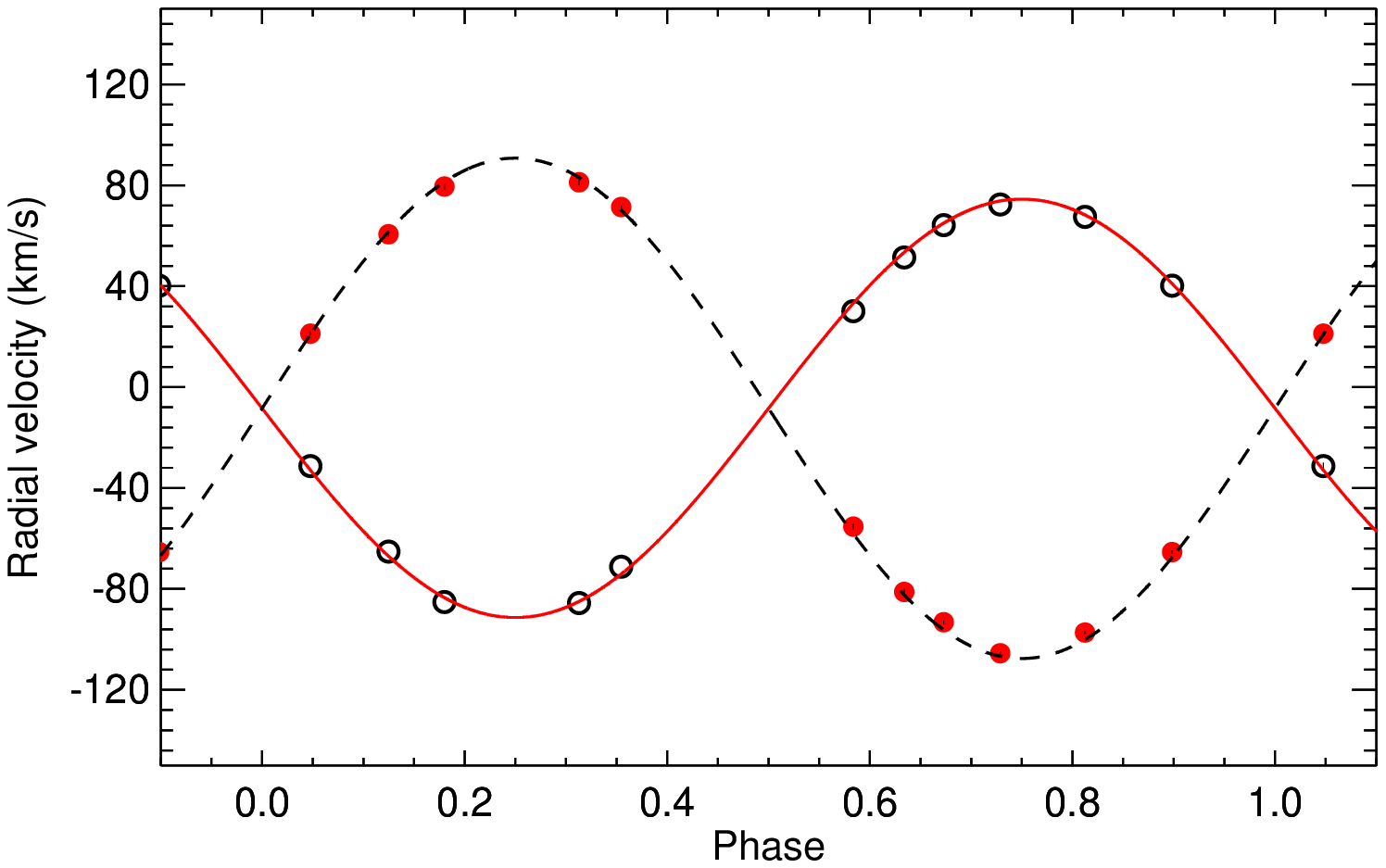}
   \caption{Radial velocities phased with the Eq.\,1 of the primary (open circles) 
   and secondary (red filled circles) component of V2653\,Oph. Error bars are 
   shown by vertical line segments, which are smaller than symbol sizes. The solid and dashed lines are 
   the computed radial velocity curves for the component stars.
   }
    \label{rv}
   \end{figure} 
\begin{figure*}
       \includegraphics[width=17cm,height=7.5cm]{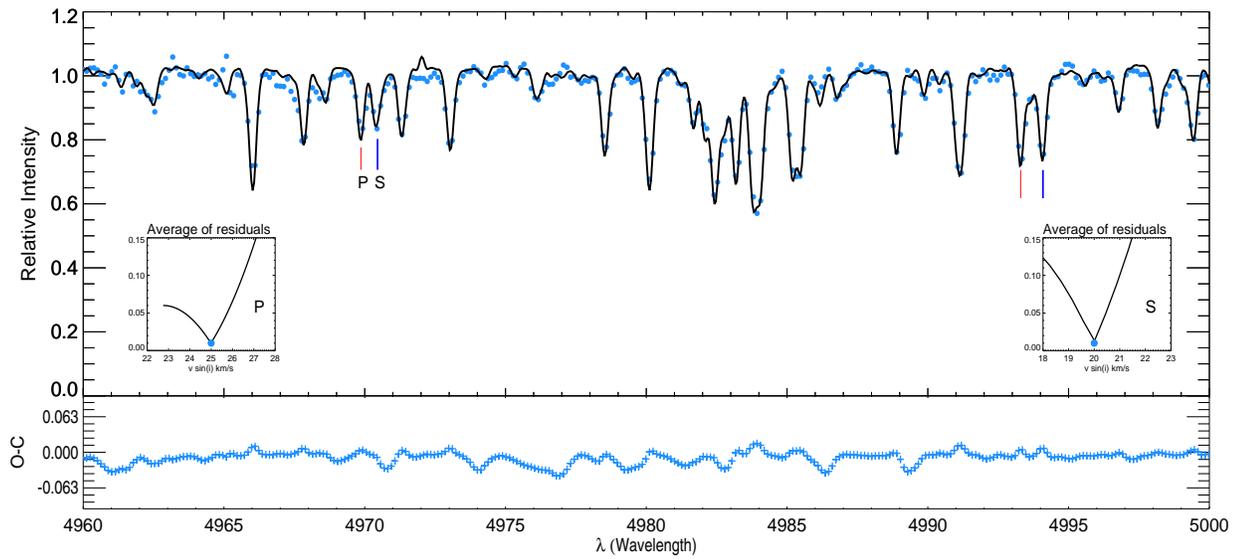}
   \caption{Comparison between the observed spectra of V2653\,Oph obtained near quadratures 
   and the best-fitting spectra around $\lambda\lambda$ 4960-5000 lines. The minimum O-C residuals
   have been obtained for the projected rotational velocities of the more massive and less massive stars
   as 25 km\,s$^{-1}$ and 20 km\,s$^{-1}$, respectively. In the bottom panel the residuals between the observed and computed 
   spectra are plotted.    
    }
    \label{spec}
   \end{figure*} 
\begin{figure}
  \centering
       \includegraphics[width=10cm,height=16cm,keepaspectratio]{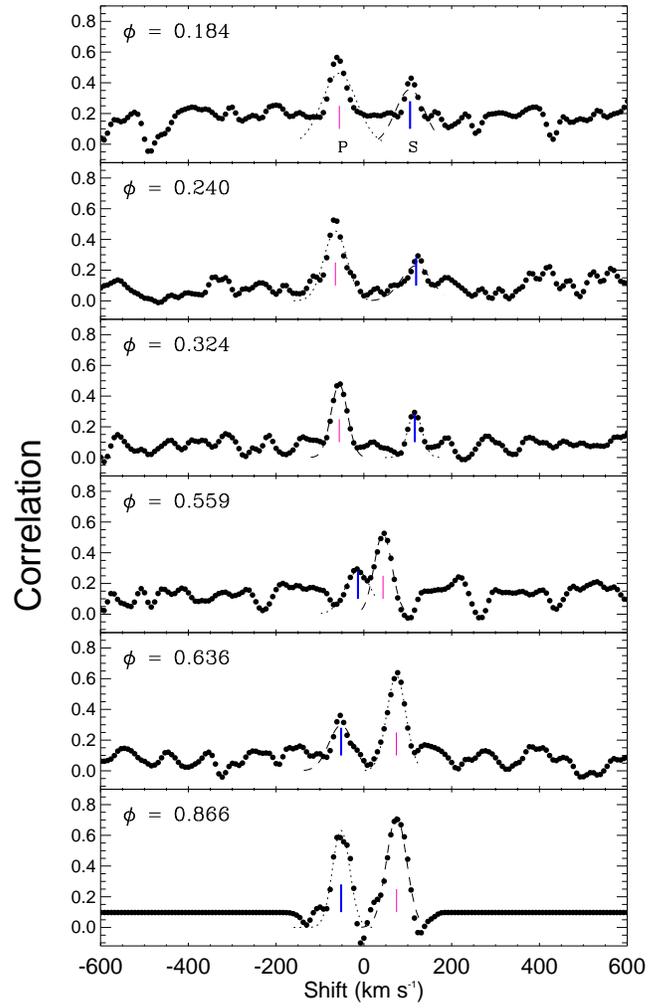}
   \caption{Sample of CCFs between V2653\,Oph and the RV template spectrum ($\iota$\,Psc) 
   around the first and second quadratures.
   }
    \label{lightrat}
   \end{figure} 
\begin{figure*}
  \centering
       \includegraphics[width=16cm,height=20cm,keepaspectratio]{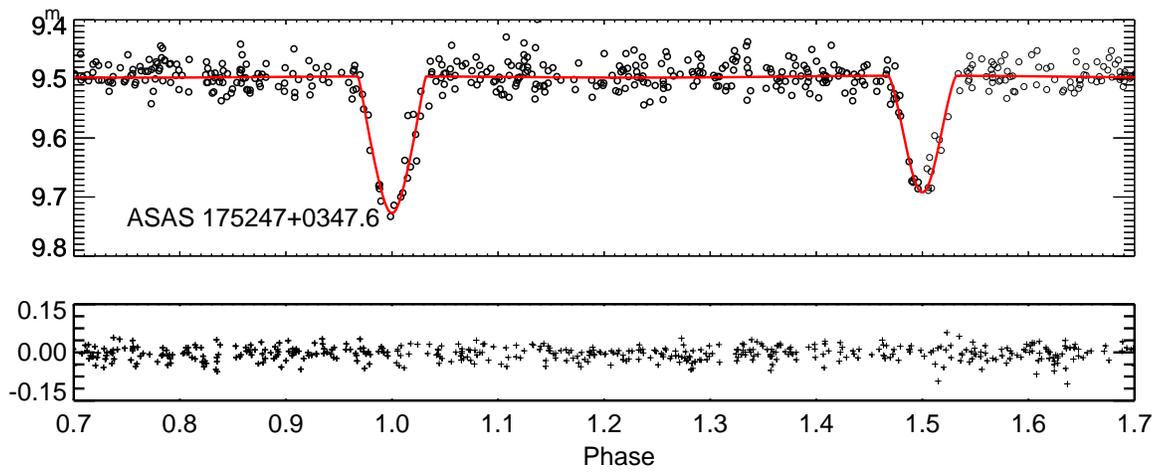}
   \caption{Full light curve of V2653\,Oph from {\em ASAS}. The WD best fit is plotted as 
blue line and the residuals of the fit are shown in the lower panel.
   }
    \label{lc}
   \end{figure*} 
 \begin{figure*}
 \centering
        \includegraphics[width=13.5cm,height=17cm,keepaspectratio]{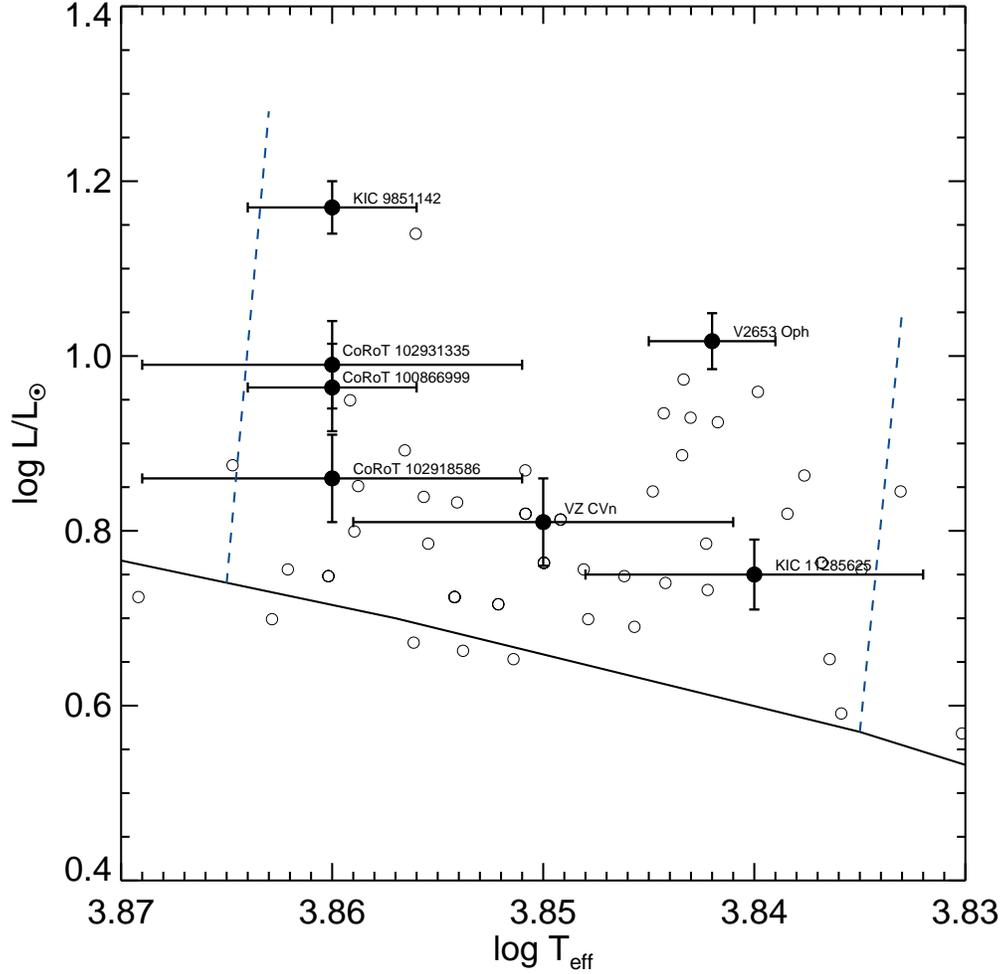}
        \caption{The domain of the $\gamma$ Dor type pulsating variables in the Hertzsprung-Russell Diagram.
         The open and filled symbols show the known $\gamma$ Dor type pulsating single stars taken from the
          catalog of \citet{henry} and the primaries of eclipsing binaries collected from individual papers,
          respectively. The  uncertainties for the $\gamma$ Dor components in eclipsing
         binaries are also shown. The dashed lines show
         the theoretical edges of the $\gamma$ Dor instability strip by \citet{warner} and the continuous line
         corresponds to the zero-age main-sequence.
        }
        \end{figure*}
\begin{figure}
  \centering
       \includegraphics[width=8.5cm,height=13cm]{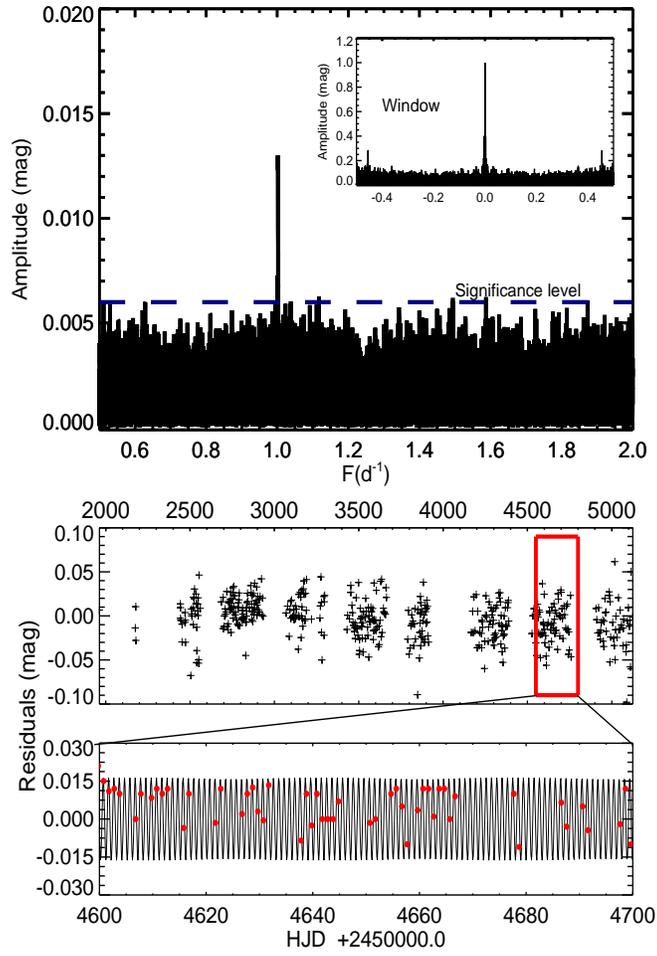}
   \caption{Periodogram of the detected frequency inside the $\gamma$ Dor frequency range for the 
   system. The significance level (∼ 0.006 mag) is indicated.
   }
    \label{rv}
   \end{figure} 
 \begin{figure}
 \centering
        \includegraphics[width=14cm,height=11cm,keepaspectratio]{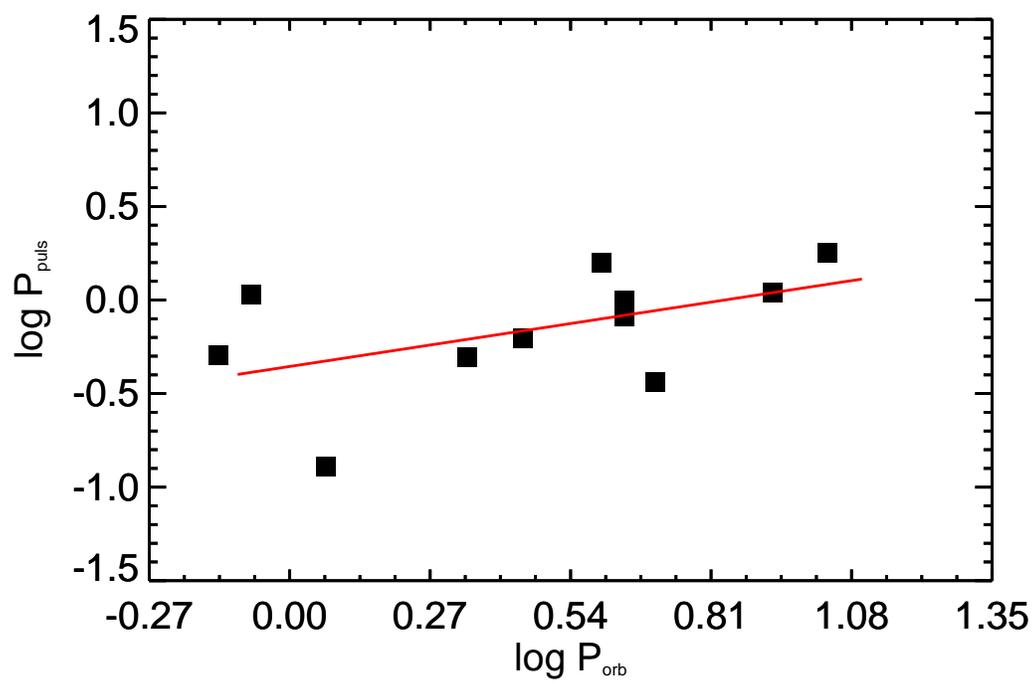}
        \caption{Relation between the pulsational and orbital periods for $\gamma$ Dor type stars in
        eclipsing binaries. The solid line represents the best fit with an equation given in the text.
        }
        \end{figure}

         \begin{figure}
 \centering
        \includegraphics[width=14cm,height=11cm,keepaspectratio]{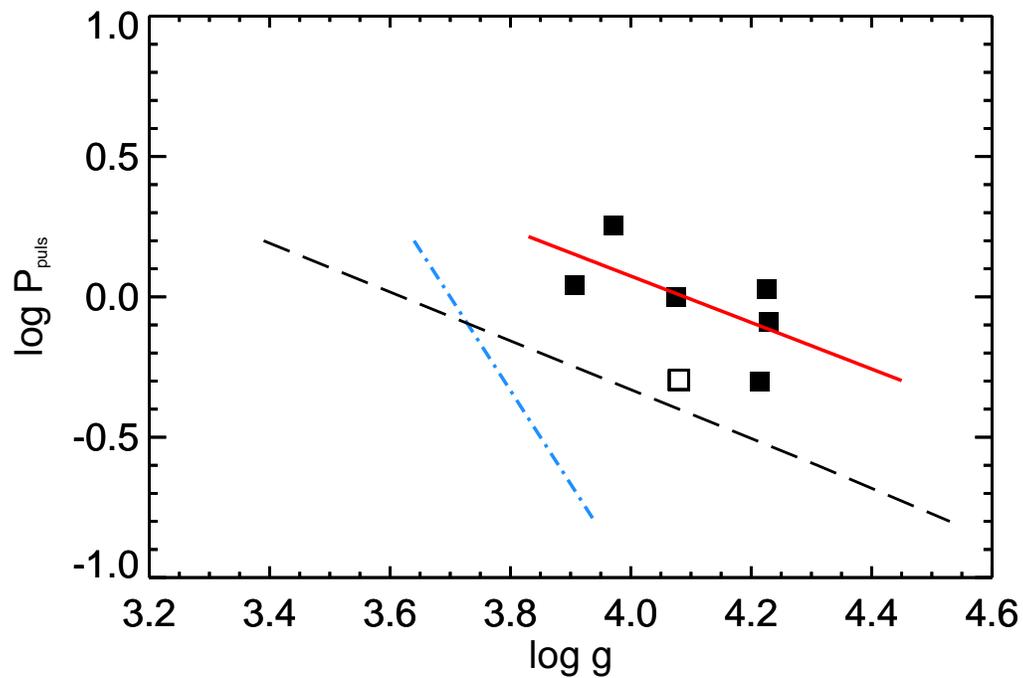}
        \caption{Plot of the pulsational periods versus the gravity of the pulsating stars (filled squares).
        The solid line represents the best fit between two parameters. The open square symbol shows V404\,Lyr for
        which no spectroscopic observation exists.  The relation, $P_{puls}$ - $g$ for
        the $\delta$-Sct stars (dot-dashed line) \citep{lia}, and for radially pulsating
        variable stars (dashed line) \citep{fer} are also shown for comparison. The gravities for radially
        pulsating stars were shifted one dex for a better comparison.
        }
        \end{figure}
        
       \begin{figure}
 \centering
        \includegraphics[width=14cm,height=11cm,keepaspectratio]{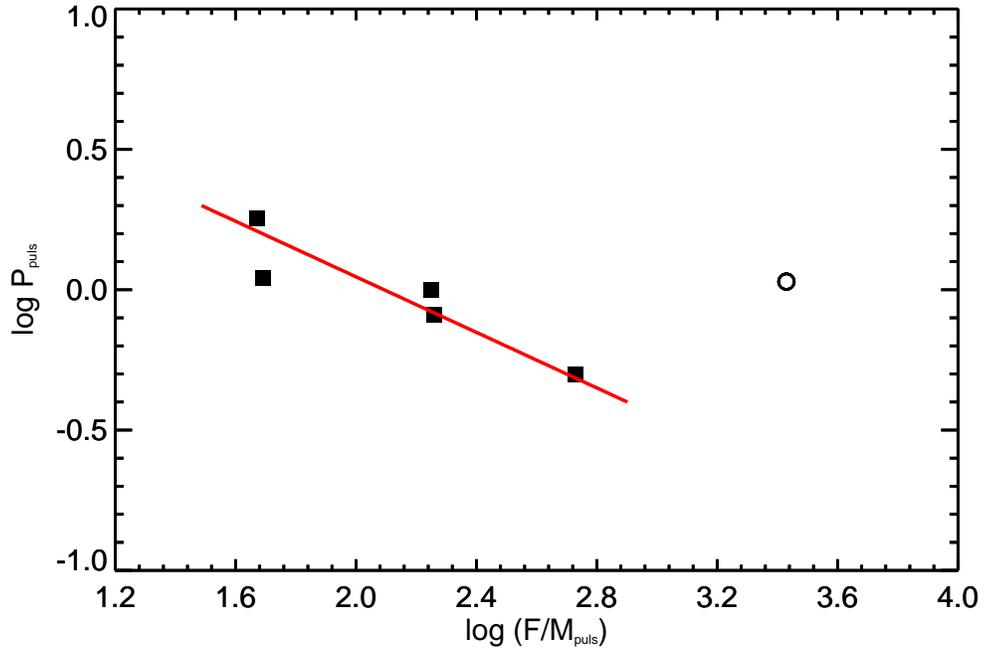}
        \caption{Relation between the pulsational period and the force on the surface of the pulsating star due to 
        the companion per unit mass of the pulsating star for $\gamma$ Dor type stars in eclipsing binaries. The system
         VZ CVn (shown by empty circle) is considerably distinguished  from the other five systems(see text). The solid line
         represents the best fit with an equation given in the text, excluding VZ CVn.
        }
        \end{figure}     

\end{document}